\documentclass[preprint,showpacs,preprintnumbers,amsmath,amssymb]{revtex4}

\usepackage{graphicx}
\usepackage{dcolumn}
\usepackage{bm}
\usepackage{amssymb}

\begin{document}
\newcommand{\vecx}{\mbox{\boldmath $x$}}
\newcommand{\vecp}{\mbox{\boldmath $p$}}

\title{
Nonextensive thermodynamics of the two-site Hubbard model: 
Canonical ensembles
}

\author{Hideo Hasegawa
\footnote{hideohasegawa@goo.jp}
}
\affiliation{Department of Physics, Tokyo Gakugei University,  
Koganei, Tokyo 184-8501, Japan}%

\date{\today}

\begin{abstract}
Canonical ensembles consisting of $M$-unit $Hubbard$ $dimers$ 
have been studies within the nonextensive statistics (NES).
The temperature dependences of the energy, entropy, specific heat and
susceptibility have been calculated for the number of
dimers, $M = 1, 2, 3$ and $\infty$.
We have assumed the relation 
between the entropic index $q$ and the cluster size $N$
given by $q=1+2/N$ ($N = 2\:M$ for $M$ dimers),
which was previously derived by several methods.
For relating the physical temperature $T$ to the Lagrange
multiplier $\beta$, two methods have been adopted:
$T=1/k_B \beta$ in the method A 
[Tsallis {\it et al.} Physica A {\bf 261}, 534 (1998)], 
and  $T=c_q/k_B \beta$ in the method B
[Abe {\it et al.} Phys. Lett. A {\bf 281}, 126 (2001)],
where $k_B$ denotes the Boltzman
constant, $c_q= \sum_i p_i^q$, and $p_i$ the 
probability distribution of the $i$th state.
The susceptibility and specific heat 
of spin dimers ({\it Heisenberg dimers})
described by the Heisenberg model
have been discussed also by using the NES
with the methods A and B.
A comparison between the two methods suggests that the method B 
may be more reasonable than the method A for nonextensive systems.
\end{abstract}

        

\maketitle
\newpage

\section{Introduction}

In the last several years,
much study has been made 
with the use of nonextensive statistics (NES)
which was initiated by Tsallis \cite{Tsallis88,Tsallis98,Tsallis04}.
When the physical quantity $Q$ of a system consisting of 
$N$ particles
is expressed by $Q\: \propto N^{\gamma}$, it is called
intensive for $\gamma=0$,
extensive for $\gamma=1$, and nonextensive  
for $\gamma \neq 1$ and $\gamma \neq 0$.
For example, in a spatially homogenous $d$-dimensional classical gas with
the attractive interaction decaying as $r^{-\alpha}$, we get
$\gamma= 2-\alpha/d$ for $0 \leq \alpha/d < 1$ (nonextensive)
and $\gamma=1$ for $\alpha/d > 1$ (extensive) \cite{Tsallis04}.
The nonextensivity is generally realized when the range of interactions
is long enough compared to the linear size of the system.
Then small-scale systems may be nonextensive 
even when the interaction is not long-ranged one.

Tsallis has proposed the NES entropy
given by \cite{Tsallis88}
\begin{equation}
S_q=k_B \left( \frac{\sum_i p_i^q -1}{1-q} \right),
\end{equation}
where $k_B$ is the Boltzman constant,
$q$ the entropic index,
and $p_i$ the probability density of the {\it i}th state.
Note that the entropy of BGS is obtained from Eq. (1)
in the limit of $q=1$.
The quantity $\mid q-1 \mid$ expresses the measure of
the nonextensivity.
The NES has been successfully
applied to a wide range of nonextensive systems
including physics, chemistry, mathematics, astronomy, 
geophysics, biology, medicine, economics, 
engineering, linguistics, and others \cite{NES}.

In our previous papers 
\cite{Hasegawa05,Hasegawa05b},
we have applied the NES to {\it Hubbard dimers}
described by the two-site Hubbard model.
The Hubbard model is
one of the most important models in solid-state physics
(for a recent review, see Ref. \cite{Kakehashi04}).
The Hubbard model consists of the tight-binding term
expressing electron hoppings and the short-range interaction
term between two electrons with opposite spins.
The Hubbard model provides us with good
qualitative description for many interesting phenomena
such as magnetism, electron correlation, and superconductivity.
In particular, the Hubbard model has been widely employed 
for a study on transition-metal magnetism. 
Thermodynamical properties of 
grand-canonical ensembles of a single Hubbard dimer 
have been calculated within the NES \cite{Hasegawa05}.
It has been shown that specific  heat and susceptibility
calculated by the NES 
may be significantly different from those calculated
by the Boltzman-Gibbs statistics 
(BGS) when the entropic index $q$ 
departs from unity, the NES with $q=1$ reducing to the BGS. 
It is interesting to compare the calculated results
with experimental data. However, experimental data for
a single dimer as adopted in Ref. \cite{Hasegawa05}, is 
not available in actual experiments.
Usually experiments on nanosystems are performed
on samples which include many clusters 
consisting of, for example, multiples dimers 
(for reviews, see Refs. 8-10). 
Iron $S=5/2$ dimers (Fe2) in $[{\rm Fe(OMe)}(dbm)_2]_2$ 
\cite{Fe2}
have the nonmagnetic, singlet ground state and their thermodynamical
property has been analyzed with the use of
the Heisenberg model \cite{Mentrup99}-\cite{Dai03}.
Similar analysis has been made for transition-metal dimers of
V2 \cite{V2}, Cr2 \cite{Cr2},
Co2 \cite{Co2}, Ni2 \cite{Ni2} and Cu2 \cite{Cu2}.
Some charge-transfer salts
like tetracyanoquinodimethan (TCNQ)
with dimerized structures, have been analyzed
by using the two-site Hubbard model within the BGS \cite{Bernstein74}.
Their susceptibility and specific heat 
were studied by taking into account
the interdimer hopping, whose effect is negligibly 
small \cite{Bernstein74}.
Such procedure may be justified within the BGS where
the specific heat and susceptibility are treated as
the extensive quantities: macroscopic measurements
are expected to reflect the property of 
a constituting dimer.
This is, however, not the case in the NES.

The purpose of the present paper is two folds.

\noindent
(1) It is interesting and indispensable to investigate
how thermodynamical properties may change when 
the size of a given cluster is varied within the NES.
It has been shown by several methods 
that the entropic $q$ of a nanosystem consisting of
independent $N$ particles is given by
\cite{Wilk00}-\cite{Raja04}
\begin{equation}
q =1+\frac{2}{N}.
\end{equation}
Bearing in mind a magnetic cluster containing 
$M$ transition-metal dimers, 
we have employed the Hubbard model to
perform NES calculations for various $M$, 
assuming the relation given by 
\begin{equation}
q =1+\frac{1}{M},
\end{equation}
which is derived from Eq. (2) with $N=2 M$ 
for $M$ dimers. We have adopted
$1.0 \leq q \leq 2.0 $ for $1 \leq M < \infty $
in this paper,
preliminary results of
the $M$ dependence of thermodynamical
quantities of clusters having been reported 
in Ref. \cite{Hasegawa05b}.

\noindent
(2) It is not clear in the current NES
how to relate the physical temperature $T$  
to the Lagrange multiplier $\beta$ \cite{Hasegawa05}.
The following two methods have been so far proposed: 
\begin{eqnarray}
T &=& \frac{1}{k_B \beta},
\hspace{2cm}\mbox{(method A)}\\
&=& \frac{c_q}{k_B \beta},
\hspace{2cm}\mbox{(method B)}
\end{eqnarray}
where $c_q=\sum_i p_i^q$ \cite{Tsallis98}.
The method A proposed in Ref. \cite{Tsallis98} 
is the same 
as the BGS.
The method B is introduced so as to satisfy the {\it zero}th law
of thermodynamical principles and the generalized Legendre
transformations \cite{Abe01}.
It has been demonstrated that the negative specific heat 
of a classical gas model which is realized in the method A \cite{Abe99},
is remedied in the method B \cite{Abe01}.
The specific heat and susceptibility calculated
by the two methods A and B are qualitatively
similar, but quantitatively different:
the nonextensivity calculated by the method A is generally
more significant than that 
calculated by the method B  \cite{Hasegawa05,Hasegawa05b}. 
In particular,
the Curie constant of the Hubbard model in the limit
of vanishing couplings 
calculated by the method A becomes spuriously large \cite{Mar00}
while that calculated
by the method B is reasonable \cite{Hasegawa05}. 
This is consistent with the result for localized free spins
\cite{Hasegawa05}. 
In order to get more insight to the unsettled issue
on the $T-\beta$ relation,
we have again made calculations with 
the use of methods A and B, by changing $M$,
which is supplementary to Ref. \cite{Hasegawa05}.

The paper is organized as follows. After briefly reviewing
the NES for the quantum system, we have derived, in Sec. 2,
expressions for the specific heat and susceptibility of
canonical ensembles of a cluster including dimers 
both in the BGS and NES, their expressions for
grand-canonical ensembles having been given 
in Ref. \cite{Hasegawa05}. 
In Sec. 3 numerical calculations of thermodynamical quantities
are reported for various values of $q$ and $M$.
The final Sec. 4 is devoted to discussion and conclusion.  
In the Appendix, the NES has been applied 
to a cluster of {\it Heisenberg dimers}.

\section{Nonextensive statistics of Hubbard dimers}

\subsection{Entropy and energy}

We have adopted nanoclusters including 
independent $M$ Hubbard dimers with negligible
interdimer interactions,
each dimer being described by the two-site Hubbard model.
The total Hamiltonian is given by
\begin{eqnarray}
H &=& \sum_{\ell=1}^M \: H_{\ell}^{(d)}, \\
H_{\ell}^{(d)} &=& -t \sum_{\sigma}
( a_{1\sigma}^{\dagger} a_{2\sigma} 
+  a_{2\sigma}^{\dagger} a_{1\sigma}) 
+ U \sum_{j=1}^2 n_{j \uparrow} n_{j \downarrow }
- \mu_B B \sum_{j=1}^2 (n_{j\uparrow} - n_{j \downarrow}), \nonumber\\
&& \hspace{10cm} \mbox{($1, 2 \in \ell$)}
\end{eqnarray}
where $H_{\ell}^{(d)}$ denotes the Hamiltonian for the $\ell$th dimer,
$n_{j\sigma}
= a_{j\sigma}^{\dagger} a_{j\sigma}$,
$a_{j\sigma}$ expresses the annihilation operator of an electron with
spin $\sigma$ on a site $j$ ($\in \ell$), 
$t$ the hopping integral,
$U$ the intraatomic interaction,
$\mu_B$ the Bohr magneton and 
$B$ an applied magnetic field.
Six eigenvalues of $H_{\ell}^{(d)}$ are given by
\begin{equation}
\epsilon_i=0, \; 2 \mu_B B, \; -2 \mu_B B, 
\; U, \; \frac{U}{2}+\Delta, \; \frac{U}{2}-\Delta,
\hspace{1cm}\mbox{for $i=1-6$}
\end{equation}
where $\Delta=\sqrt{U^2/4+4 t^2}$ 
\cite{Bernstein74}\cite{Suezaki72}.
The number of eigenstates of the total Hamiltonian 
$H$ is $6^M$.

First we employ the BGS, in which
the canonical partition function for $H$ 
is given by \cite{Bernstein74}\cite{Suezaki72}
\begin{eqnarray}
Z_{BG}&=& {\rm Tr} \: {\rm exp}(- \beta H), \\
&=& \sum_{i_1=1}^6 \cdot \cdot \sum_{i_M=1}^6
{\rm exp}[-\beta (\epsilon_{i_1} + \cdot \cdot+ \epsilon_{i_M})], \\
&=& [Z_{BG}^{(d)}]^M, \\
Z_{BG}^{(d)}&=&1+2 \:{\rm cosh}(2 \beta \mu_B B) + e^{- \beta U}
+ 2 \:e^{- \beta U/2} {\rm cosh} (\beta \Delta),
\end{eqnarray}
where $\beta=1/k_B T$, Tr denotes the trace and 
$Z_{BG}^{(d)}$ the partition function
for a single dimer.
By using the standard method in the BGS, 
we can obtain various thermodynamical quantities of the system 
\cite{Bernstein74,Suezaki72,Shiba72}.
Because of the product expression given by Eq. (11),
the energy and entropy are proportional to $M$:
$E_{BG}= M E_{BG}^{(d)}$ and $S_{BG}= M S_{BG}^{(d)}$
where $E_{BG}^{(d)}$ and $S_{BG}^{(d)}$ are
for a single dimer.
This is not the case in the NES as will be discussed below.

The entropy $S_q$ in the Tsallis NES
is defined by \cite{Tsallis88,Tsallis98}
\begin{equation}
S_q=k_B \left( \frac{{\rm Tr} \:(\rho_q^q) - 1}{1-q} \right).
\end{equation}
Here $\rho_q$ stands for 
the generalized canonical density matrix,
whose explicit form will be determined shortly [Eq. (16)].
We impose the two constraints given by 
\begin{eqnarray}
{\rm Tr} \:(\rho_q)&=&1, \\
\frac{{\rm Tr} \:(\rho_q^q H)}{{\rm Tr} \:(\rho_q^q)}
&\equiv& <H>_q = E_q,
\end{eqnarray}
where the normalized formalism is adopted \cite{Tsallis98}.  
The variational condition for the entropy with
the two constraints given by Eqs. (14) and (15)
yields
\begin{equation}
\rho_q=\frac{1}{X_q} {\rm exp}_q 
\left[ -\left( \frac{\beta}{c_q} \right) (H-E_q) \right],
\end{equation}
with
\begin{equation}
X_q={\rm Tr}\: \left( {\rm exp}_q 
\left[-\left( \frac{\beta}{c_q} \right) (H-E_q)\right] \right),
\end{equation}
\begin{equation}
c_q= {\rm Tr} \:(\rho_q^q) = X_q^{1-q},
\end{equation}
where ${\rm exp}_q [x]$
expresses the $q$-exponential function defined by
\begin{eqnarray}
{\rm exp}_q [x] &=& [1+(1-q)x]^{\frac{1}{1-q}},
\hspace{2cm}\mbox{for $(1-q)x > 0$}\nonumber \\
&=&0,
\hspace{4cm}\mbox{otherwises}
\end{eqnarray}
and $\beta$ is a Lagrange multiplier:
\begin{equation}
\beta=\frac{\partial S_q}{\partial E_q}.
\end{equation}
The trace in Eq. (17) or (18) is performed over
the $6^M$ eigenvalues, for example, as
\begin{eqnarray}
X_q &=&\sum_{i_1=1}^6 \cdot \cdot \sum_{i_M=1}^6
\left( {\rm exp}_q \left[ - \left( \frac{\beta}{c_q} \right)
(\epsilon_{i_1}+ \cdot \cdot + \epsilon_{i_M}-E_q)\right] \right),
\nonumber \\
&\ \equiv & \sum_i 
\left( {\rm exp}_q \left[ - \left( \frac{\beta}{c_q} \right)
(\epsilon_i-E_q) \right] \right),
\end{eqnarray}
where the following conventions are adopted:
\begin{eqnarray}
i &=& (i_1, \cdot \cdot i_M), \\ 
\sum_i &=& \sum_{i_1=1}^6 \cdot \cdot \sum_{i_M=1}^6, \\
\epsilon_i &=& \epsilon_{i_1}+ \cdot \cdot + \epsilon_{i_M}.
\end{eqnarray}

It is noted that in the limit of $q = 1$,
Eq. (17) reduces to
\begin{equation}
X_1 = Z_{BG} \:{\rm exp}[\beta E_1] 
=[Z_{BG}^{(d)} \: {\rm exp}\;(\beta E_{BG}^{(d)})]^M.
\end{equation}
For $q \neq 1$, however, $X_q$ cannot be expressed
as a product form because
of the property of the $q$-exponential function:
\begin{eqnarray}
{\rm exp}_q(x+y) 
&\neq& \exp_q(x)\;\exp_q(y) 
\hspace{1cm}\mbox{for $q \neq 1$}
\end{eqnarray}

It is necessary to point out that $E_q$ in Eq. (15) includes
$X_q$ which is expressed by $E_q$ in Eq. (17).
Then $E_q$ and $X_q$ have to be determined self-consistently
by Eqs. (15)-(19) with Eq. (4) or (5) 
for a given temperature $T$.
The calculation of thermodynamical quantities
in the NES generally
becomes more difficult than that in BGS.

\subsection{Specific heat}
The specific heat
in the NES is given by \cite{Hasegawa05,Hasegawa05b}
\begin{equation}
C_q= \left( \frac{d \beta}{d T} \right) 
\left( \frac {d E_q}{d \beta} \right).
\end{equation}
Because $E_q$ and $X_q$ are determined by
Eqs. (15)-(19), we get simultaneous equations for
$d E_q/d \beta$ and  $d X_q/d \beta$, given by
\begin{eqnarray}
\frac {d E_q}{d \beta} 
&=& a_{11}  \left( \frac {d E_q}{d \beta} \right)
+ a_{12} \left( \frac {d X_q}{d \beta} \right) + b_1, \\
\frac {d X_q}{d \beta} 
&=& a_{21}  \left( \frac {d E_q}{d \beta} \right)
+ a_{22} \left( \frac {d X_q}{d \beta} \right),
\end{eqnarray}
with
\begin{eqnarray}
a_{11}&=& q \beta X_q^{q-2} 
\sum_i w_i^{2q-1} \epsilon_i,  \\
a_{12}&=& -X_q^{-1} E_q
-\beta q (q-1) X_q^{q-3} \sum_i w_i^{2q-1}
\epsilon_i (\epsilon_i -E_q), \\
a_{21}&=& \beta X_q^q,\\
a_{22}&=& 0,\\
b_1&=& - q X_q^{q-2} \sum_i w_i^{2q-1}
\epsilon_i (\epsilon_i-E_q),\\
w_i &=& <i\mid {\rm exp}_q 
\left [- \left( \frac{\beta}{c_q}\right) 
(H- E_q) \right] \mid i>, \nonumber \\
&=& \left[1-(1-q) \left( \frac{\beta}{c_q} \right)
(\epsilon_i - E_q) \right]^{\frac{1}{1-q}},\\
X_q &=& \sum_i w_i.
\end{eqnarray}
The specific heat is then given by
\begin{equation}
C_q= \left( \frac{d \beta}{d T} \right) 
\left( \frac{b_1}{1-a_{11}-a_{12}a_{21}} \right).
\end{equation}
with
\begin{eqnarray}
\frac{\partial \beta}{\partial T}
&=& - \beta^2, 
\hspace{7cm} \mbox{(method A)} \\
&=& - \left( \frac{\beta^2}
{X_q^{1-q} - \beta (1-q) X_q^{-q} 
(d X_q/d \beta) }\right), 
\hspace{1cm} \mbox{(method B)} 
%
%
\end{eqnarray}
In the limit of $q \rightarrow 1$,
Eqs. (27)-(39) yield the specific heat
in the BGS, given by \cite{Hasegawa05}
\begin{equation}
C_{BG} = \frac{d E_{BG}}{d T}
= k_B \beta^2 (<\epsilon_i^2>_1 - <\epsilon_i>_1^2),
\end{equation}
where $<\cdot>_1$ is defined by Eq. (15).

\subsection{Magnetization}

We discuss the field-dependent magnetization, which is given
in the BGS, by
\begin{eqnarray}
m_{BG}&=& - \frac{\partial 
F_{BG}}{\partial B} 
=\left < \mu_i \right>_{1},\\
&=& \frac{4 \mu_B \:{\rm sinh}(2\beta B)}{Z_{BG}}
\end{eqnarray}
where $\mu_i=- \partial \epsilon_i/\partial B$, 
$Z_{BG}$ and $<\cdot >_{1}$ are
given by Eqs. (9) and (15), respectively.

In the NES,
the magnetization $m_q$ is given by 
\begin{eqnarray}
m_q&=& -\frac{\partial E_q}{\partial B} 
+ (k_B \:\beta)^{-1} \frac{\partial S_q}{\partial B},\\
&=& -\frac{\partial E_q}{\partial B} 
+ \beta^{-1} X_q^{-q} \frac{\partial X_q}{\partial B}
\end{eqnarray}
By using Eqs. (15)-(19),
we get the simultaneous equations for 
$\partial E_q/\partial B$ and 
$\partial X_q/\partial B$ given by
\begin{eqnarray}
\frac{\partial E_q}{\partial B}
&=&a_{11} \left( \frac{\partial E_q}{\partial B} \right)
+ a_{12} \left( \frac{\partial X_q}{\partial B} \right) + d_1, \\
\frac{\partial X_q}{\partial B}
&=&a_{21} \left( \frac{\partial E_q}{\partial B} \right)
+ a_{22} \left( \frac{\partial X_q}{\partial B} \right) + d_2, 
\end{eqnarray}
with
\begin{eqnarray}
d_1&=& - X_q^{-1} \sum_i w_i^q \mu_i
+ \beta q X_q^{q-2} \sum_i w_i^{2q-1} \epsilon_i
\mu_i,\\
d_2&=& \beta X_q^{q-1} \sum_i w_i^q \mu_i,
\end{eqnarray}
where $a_{ij}$ ($i,j=1,2$) are given by Eqs. (30)-(33).
We obtain $m_q$ given by
\begin{eqnarray}
m_q&=&\left( \frac{-c_{12}+\beta^{-1}X_q^{-q}(1-c_{11})}
{1-c_{11}-c_{12}c_{21}} \right) d_2, \\
&=& X_q^{-1} \sum_i \:w_i^q \:\mu_i
= <\mu_i>_q. 
\end{eqnarray}
In the limit of $q \rightarrow 1$,
Eqs. (47) and (48) reduce to
\begin{eqnarray}
d_1&=& - \left< \mu_i \right>_1
+ \beta \left< \epsilon_i \mu_i \right>_1,\\
d_2&=& \beta X_1 \left< \mu_i \right>_1,
\end{eqnarray}
where $<\cdot>_1$ is given by Eq. (15) with $q=1$.
By using Eq. (50), we get 
$m_1 = m_{BG}$, which is given by Eq. (42).

\subsection{Susceptibility}

The high-field susceptibility 
in the NES is given by
\begin{equation}
\chi_q(B)= \frac{\partial m_q}{\partial B}.
\end{equation}
The zero-field susceptibility 
$\chi_q(B=0)$ is given by
\cite{Hasegawa05,Hasegawa05b}
\begin{eqnarray}
\chi_q&=& \chi_q(B=0)
= -E_q^{(2)} + \beta^{-1} X_q^{-q}X_q^{(2)},
\end{eqnarray}
where 
$E_q^{(2)}=\partial^2 E_q/\partial B^2\mid_{B=0}$ and 
$X_q^{(2)}=\partial^2 X_q/\partial B^2\mid_{B=0}$.
With the use of Eqs. (15)-(19), we get simultaneous equations
for $E_q^{(2)}$ and $X_q^{(2)}$ given by
\begin{eqnarray}
E_q^{(2)}&=&a_{11} E_q^{(2)}+ a_{12} X_q^{(2)} + f_1, \\
X_q^{(2)}&=&a_{21} E_q^{(2)}+ a_{22} X_q^{(2)} + f_2, 
\end{eqnarray}
with
\begin{eqnarray}
f_1&=& -2 \:\beta \:q \:X_q^{q-2} \sum_i w_i^{2q-1}\: \mu_i^2,\\
f_2&=& \beta^2 \:q \:X_q^{2(q-1)} \sum_i w_i^{2q-1} \:\mu_i^2,
\end{eqnarray}
where $a_{ij}$ ($i,j=1,2$) are given by Eqs. (30)-(33).
From Eqs. (54)-(58), we get
\begin{eqnarray}
\chi_q&=& \frac{f_2}{a_{21}}
=\beta q X_q^{q-2} \sum_i w_i^{2q-1} 
\mu_i^2\mid_{B=0}.
\end{eqnarray}
In the limit of $q=1$, Eq. (59) yields
the susceptibility in BGS:
\begin{eqnarray}
\chi_{BG} &=& \beta
<\mu_i^2 \mid_{B=0}>_1 ,\\
&=& \left(\frac{\mu_B^2}{k_B T}\right)
\left( \frac{8}{3+e^{-\beta U}+2 e ^{-\beta U/2}
\:{\rm cosh}(\beta \Delta)} \right).
\end{eqnarray}

\section{Calculated results}

\subsection{The $q$ dependence}

We have performed numerical calculations by changing 
the index $q$ or the size of a cluster $M$ in the NES. 
Simultaneous equations for $E_q$ and $X_q$
given by Eqs. (15)-(19) have been solved
by using the Newton-Raphson method 
with initial values of $E_1$ and $X_1$
obtained from BGS ($q=1$). 
The magnetic field $B$ in Eq. (8) is set zero in calculating
the entropy, energy and specific heat.
The calculated quantities are given {\it per dimer}. 

\vspace{0.5cm}
\begin{center}
{$------- Fig. 1 ---------$}
\end{center}
\vspace{0.5cm}

First we treat the entropic index $q$ as a free parameter
for a single dimer.
Figures 1(a)-1(f) show the temperature dependence of 
the energy $E_q$ calculated for $B=0$.
Bold solid curves in Fig. 1(a), 1(b) and 1(c) show
$E_1$ in the BGS
calculated for $U/t=0$, 5 and 10, respectively.
The ground-state energy at $T=0$ is $E_1/t=$ -2.0, -0.70156 and -0.38516
for $U/t=0$, 5 and 10, respectively.
With increasing $q$ value above unity, the gradient
of $E_q$ is much decreased in the method A, as shown in Figs. 1(a)-1(c).
This trend is, however, much reduced in the method B, as Figs. 1(d)-1(f)
show. 
This behavior is more clearly seen in the temperature dependence
of the specific heat $C_q$, as will be discussed shortly
[Figs. 3(a)-3(f)].

\vspace{0.5cm}
\begin{center}
{$------- Fig. 2 ---------$}
\end{center}
\vspace{0.5cm}

Temperature dependences of the entropy 
for $B=0$ are plotted in Figs. 2(a)-2(f).
Figures 2(a), 2(b) and 2(c) express $S_q$
for $U/t=0$, 5 and 10, respectively,
calculated by the method A, and Figs. 2(d)-2(f)
those calculated by the method B.
Bold curves denote the results for the BGS, where the entropy 
is quickly increased at low temperature when the interaction is increased.
When the $q$ value is more increased above unity, $S_q$ 
is more rapidly increased
at very low temperatures and its saturation value 
at higher temperatures becomes smaller.
This behavior is commonly realized in the results calculated
by the methods A and B.
A difference between the two results is ostensibly small
because $S_q$ shows a saturation at low temperatures.

\vspace{0.5cm}
\begin{center}
{$------- Fig. 3 ---------$}
\end{center}
\vspace{0.5cm}

Figures 3(a)-3(f) show the specific heat calculated
for $B=0$. 
$C_1$ in BGS for $U/t=0$ shown by the bold solid curve 
in Fig. 3(a), has a peak at $k_B T/t \sim 0.65$.
Figure 3(c) shows that
for $U/t=10$, this peak splits into two.
A lower peak arises from low-lying collective
spin-wave-like excitations while higher one
from single-particle excitations \cite{Bernstein74,Shiba72}.
For intermediate $U/t=5$ these two peaks overlap [Fig. 3(b)]. 
The temperature
dependences of the specific heat $C_q$
calculated with the use of the method A
for $U/t=0$, 5 and 10 
are plotted in Figs. 3(a), 3(b) and 3(c), respectively.
We note that when $q$ is larger than unity, peaks become broader.
Figures 3(d), 3(e) and 3(f) show the temperatures
dependence of the specific heat $C_q$ 
calculated by the method B
for $U/t=0$, 5 and 10, respectively.
Although general property of the $q$ dependence of 
the specific heat of the method B is similar 
to that of the method A, the effect of the nonextensivity
in the method B becomes smaller than that in the method A.

\vspace{0.5cm}
\begin{center}
{$------- Fig. 4 ---------$}
\end{center}
\vspace{0.5cm}

The BGS susceptibility for $U/t=0$ has a peak at
$k_B T/t \sim 0.65$ as Fig. 4(a) shows.
With increasing $U/t$, the magnitude of $\chi_{BG}$
is enhanced by the interaction,
and its peak position becomes lower \cite{Bernstein74,Shiba72},
as Figs. 4(b) and 4(c) show:
the horizontal scale of Fig. 4(c) is
enlarged compared with Figs. 4(a) and 4(b).
The temperature dependences of the susceptibility $\chi_q$
calculated by the method A for $U/t=0$, 5 and 10 
are plotted in Figs. 4(a), 4(b) and 4(c), respectively.
We note that as increasing $q$ above unity, the peak
in $\chi_q$ becomes broader.
Figures 4(d), 4(e) and 4(f) show the temperature
dependence of the susceptibility $\chi_q$
calculated by the method B 
for $U/t=0$, 5 and 10, respectively.
Again the effect of the nonextensivity in the
method B becomes smaller than that in the method A.

\subsection{The $M$ dependence}

\vspace{0.5cm}
\begin{center}
{$------- Fig. 5 ---------$}
\end{center}
\vspace{0.5cm}

In order to study how thermodynamical quantities of a cluster
with Hubbard dimers depend on its size $M$, we have made 
NES calculations, assuming the $q$ value
for a given $M$ value with
the $M-q$ relation given by Eq. (3).
Results for $M=\infty$ correspond to 
those of the BGS ($q=1$).
Figures 5(a)-5(d) show the results for non-interacting
case of $U/t=0$. 
The specific heat and susceptibility
shown in Figs. 5(a) and 5(b), have been
calculated by the method A 
with $q=$ 2.0, 1.5, and 1.333 for
$M=1$, 2 and 3, respectively.
Figures 5(c) and 5(d) express 
$C_q$ and $\chi_q$, respectively, calculated by the method B.
We note that
physical quantities in a small cluster with $M \sim 1-3$ are 
rather different from
those of bulk-like systems with $M=\infty$,
although properties of clusters gradually approach
those of bulk with increasing $M$.

\vspace{0.5cm}
\begin{center}
{$------- Fig. 6 ---------$}
\end{center}
\vspace{0.5cm}

Similar results for finite interaction of $U/t=5$
are shown in Figs. 6(a)-6(d).
The specific heat and susceptibility
plotted in Figs. 6(a) and 6(b), respectively, have been
calculated by the method A
for $M=1$, 2, 3 and $\infty$.
Figures 6(c) and 6(d) show
$C_q$ and $\chi_q$, respectively, 
calculated by the method B for $U/t=5$.
The results for small $M$ are very different from those
for $M = \infty$.
We note that the $M$ dependences of
$C_q$ and $\chi_q$ of Hubbard dimers shown in 
Figs. 6(a)-6(d) are similar to those of spin dimers described by 
the Heisenberg model [Figs. 13(a)-13(d)],
details being discussed in the Appendix A.
This is not surprising because the Hubbard model
with the strong coupling and the half-filled electron occupancy,
reduces to the Heisenberg model with the antiferromagnetic
exchange interaction.

\vspace{0.5cm}
\begin{center}
{$------- Fig. 7 ---------$}
\end{center}
\vspace{0.5cm}

We have calculated the $M$ dependence of the maximum
values of $C^*_q$ and $\chi^*_q$ and corresponding
temperatures of $T^*_C$ and $T^*_{\chi}$.
Figure 7(a) shows $T^*_C$ and $T^*_{\chi}$, 
and Fig. 7(b) depicts $C^*_q$ and $\chi^*_q$,
all of which are plotted against $1/M$:
solid and dashed lines denote results 
calculated by the methods A and B, respectively.
It is shown in Fig. 7(a) that with increasing $1/M$,
$T^*_{\chi}$ calculated by the method A
is much increased than that calculated by the method B.
We note also that with increasing $1/M$,
$T^*_{C}$ of the method B is increased
while that of the method A is decreased.
Figure 7(b) shows that $C_q^*$ in the method A is smaller
than that in the method B, whereas
$\chi^*_q$ in the method A is the 
same as that in the method B.

\section{Discussions and Conclusions}

\vspace{0.5cm}
\begin{center}
{$------- Fig. 8 ---------$}
\end{center}
\vspace{0.5cm}

Although we have discussed the temperature dependence
of physical quantities in the preceding section,
it is worthwhile to study their magnetic-field dependence.
Figures 8(a), 8(b) and 8(c) show the magnetization $m_q$ 
as a function of the magnetic field $B$ 
for $U/t=0$, 1 and 10, respectively, at $k_B T/t=1.0$
calculated by the method A: results calculated by
the method B is not so different from them [Figs. 10(a)].
When $q$ is increased above unity, the magnetization
at lower fields ($h/t < 1$) is decreased whereas
at higher fields ($h/t > 1$) it is much increased.
This is consistent with the calculation of the
susceptibility shown in Fig. 3(a)-3(c),
where $\chi_q$ at $k_B T/t=1.0$ is smaller for larger $q$:
the susceptibility stands for the initial gradient of $m_q$
at $B=0$.

\vspace{0.5cm}
\begin{center}
{$------- Fig. 9 ---------$}
\end{center}
\vspace{0.5cm}

\vspace{0.5cm}
\begin{center}
{$------- Fig. 10 ---------$}
\end{center}
\vspace{0.5cm}

The field dependence of physical quantities
are discussed in more details.
Figure 9 shows the $B$ dependence of the six eigenvalues
of $\epsilon_i$ for $U/t=5$ [Eq. (8)].
We note the crossing of the eigenvalues
of $\epsilon_3$ and $\epsilon_6$ at the critical filed:
\begin{equation}
\mu_B B_c = \sqrt{\frac{U^2}{16}+ t^2}-\frac{U}{4},
\end{equation}
leading to $\mu_B B_c/t=0.351$ for $U/t=5$.
At $B=B_c$ the magnetization $m_q$ is rapidly increased
as shown in Figs. 10(a) and 10(b) for $k_B T/t=1.0$
and 0.1, respectively:
the transition at lower temperature is more
evident than at higher temperature. 
This level crossing also yields a peak in $\chi_q$ 
[Figs. 10(c) and 10(d)] and
a dip in $C_q$ [Figs. 10(e) and 10(f).
It is interesting that
the peak of $\chi_q$ for $q=1.5$ in the NES 
is more significant than that
in the BGS whereas that of $C_q$ of the former
is broader than that of the latter. 
When the temperature becomes higher, these peak structures
become less evident.
Similar phenomenon in the field-dependent specific heat
and susceptibility have been pointed out in the
Heisengerg model within the BGS \cite{Kuzmenko04}.

Figure 10(a) and 10(b) remind us the quantum tunneling of magnetization
observed in magnetic molecular clusters such as Mn4, 
Mn12 and Fe8 \cite{Mn12}.
It originates from the level crossing of 
magnetic molecules which are parallel and anti-parallel
to the easy axis when a magnetic field is applied.

The $N-q$ relation given by $q=1+2/N$ [Eq. (2)] 
has been derived from 
the average of the BGS partition function
of ${\rm exp}(-\beta \epsilon)$ with 
$\epsilon=\sum_i \epsilon_i$
over fluctuating $\beta$
fields, as given by \cite{Wilk00}-\cite{Raja04}
\begin{equation}
w(\{\epsilon_i\})=\int_0^{\infty} \: d\beta 
\:{\rm exp}\left( -\beta \sum_{i=1}^N \epsilon_i \right)\: f^B(\beta) 
= {\rm exp}_q \left[-\beta_0 \sum_{i=1}^N \epsilon_i \right],
\end{equation}
with 
\begin{eqnarray}
f^B(\beta)&=& \frac{1}{\Gamma \left( \frac{N}{2} \right)}
\left( \frac{N}{2\beta_0} \right)^{\frac{N}{2}}
\beta^{\frac{N}{2}-1} 
{\rm exp}\left( -\frac{N \beta}{2\beta_0} \right), \\
%
\beta_0
&=& <\beta>, \\
\frac{2}{N}&=& \frac{<\beta^2>-<\beta>^2}{<\beta>^2}.
\end{eqnarray}
Here $< Q >$ stands for the expectation value of $Q$
averaged over the $\Gamma$ (or $\chi^2$) distribution 
function $f^B(\beta)$, 
$\beta_0$ the average of the fluctuating $\beta$
and $2/N$ its variance.
The $\Gamma$ distribution is emerging from the sum
of squares of $N$ Gaussian random variables.
Alternatively, by using the large-deviation approximation,
Touchette \cite{Touchette02} has obtained the distribution
function $f^T(\beta)$, in place of $f^B(\beta)$, given by
\begin{eqnarray}
f^T(\beta)&=& \frac{\beta_0}{\Gamma 
\left( \frac{N}{2} \right)}
\left( \frac{N \beta_0}{2} \right)^{\frac{N}{2}}
\beta^{-\frac{N}{2}-2} 
{\rm exp}\left( -\frac{N \beta_0}{2\beta} \right).
\end{eqnarray}

\vspace{0.5cm}
\begin{center}
{$------- Fig. 11 ---------$}
\end{center}
\vspace{0.5cm}

Figure 11 shows the $f^B$- and $f^T$-distribution functions
for various $N$ values.
For $N \rightarrow \infty$,
both reduce to the delta-function densities, and
for a large $N=100$, both distribution functions
lead to similar results.
For a small $N \;(< 10)$, however, 
there is a clear difference between the two distribution 
functions.
We note that a change of variable $\beta \rightarrow \beta^{-1}$
in $f^T$ yields the distribution function similar to $f^B$.  
It should be noted that $f^T$ cannot lead to the
$q$-exponential function which plays a crucial role
in the NES. For a large $\epsilon$,
$f^T$ leads to the stretched exponential form of
$w(\epsilon) \sim e^{c \sqrt{\epsilon}}$
while $f^B$ yields the power form of
$w(\epsilon) \sim \epsilon^{-\frac{1}{q-1}}$.
This issue of $f^B$ vs. $f^T$ is related to the {\it superstatistics},
which is currently studied with much interest
\cite{Beck04}.

In deriving Eq. (63), we have implicitly assumed that 
the distance between sparsely distributed clusters
is larger than $\xi$, the {\it coherence} length of 
the fluctuating $\beta$ field over which
the field $\beta$ uniformly fluctuates,
and that the linear size of the clusters
is smaller than $\xi$.
If the population of constituting dimers 
in a given cluster is sparse such
that the distance between dimers is larger than $\xi$
and a local fluctuating $\beta_i$ field around a dimer $i$
is almost independent from the local $\beta_j$ field
around another dimer $j$,
Eq. (64) is replaced by
\begin{equation}
w(\{\epsilon_i\})=\Pi_{i=1}^N \int_0^{\infty} \: d\beta_i 
\:{\rm exp}(-\beta_i \epsilon_i)\: f^B(\beta_i) 
= \Pi_{i=1}^N {\rm exp}_{q'}[-\beta_0 \epsilon_i],
\end{equation}
with $q'=3$. 
Actual distribution of dimers is expected to
lie between the two extreme cases given by Eqs. (63) and (68) 
\cite{Beck02}.

Although results calculated by
the two methods A and B are qualitatively similar,
there are some quantitative difference,
as previously obtained 
in Refs. \cite{Hasegawa05} and \cite{Hasegawa05b}. 
When we calculate
the Curie constant $\Gamma_q$ of the susceptibility
defined by $\chi_q(T)=(\mu_B^2/k_B)[\Gamma_q(T)/T]$,
the ratio between $\Gamma_q^{(A)}$ and $\Gamma_q^{(B)}$
calculated by the two methods, 
is given by
\begin{eqnarray}
\frac{\Gamma_q^{(A)}}{\Gamma_q^{(B)}}
&=& c_q, \\
&=& 4^{(q-1)}, 
\hspace{2.5cm}\mbox{for $T=0$} \\
&=& 6^{(q-1)}.
\hspace{2.5cm}\mbox{for $T=\infty$}
\end{eqnarray}
In general, $\Gamma_q^{(B)}$ depends on $t$, $U$ and $T$. 
In the limit of $t=0$, for example, it is given by \cite{Note9}
\begin{eqnarray}
\Gamma_q^{(B)} &=& 2\:q, 
\hspace{2.5cm}\mbox{for $T=0$} \\
&=& \frac{4}{3}\:q.
\hspace{2.5cm}\mbox{for $T=\infty$}
\end{eqnarray}
The result of the method A given by Eqs. (69)-(73)
leads to anomalously large Curie constant
compared to that of the method B.
Equation (69) is consistent
with the result for spin dimers described
by the Heisenberg model (for details, see Appendix A)
\cite{Hasegawa05}.
A comparison between Eqs. (35) and (63) yield
the average temperature $<T>$ given by
\begin{equation}
\frac{1}{k_B <T>} \simeq \beta_o = \frac{c_q}{\beta}
\end{equation}
which supports the method B.
These results suggest that the method B is 
more appropriate than the method A.
This is consistent with recent theoretical analyses made 
in \cite{Suyari05,Wada05}
(for related discussions, see Ref. \cite{Abe05}].

In summary,
within the framework of the NES,
thermodynamical properties 
have been discussed of
a cluster including $M$ dimers, each of which is described by
the two-site Hubbard model.
We have demonstrated that the thermodynamical properties
of small-scale systems are rather different
from those of bulk systems.
Owing to recent progress in atomic engineering, 
it is possible
to synthesize molecules containing relatively small numbers of 
magnetic atoms with the use of various methods
(for reviews, see refs. 8-10). 
Small-size magnetic systems ranging from grains (micros),
nanosystems, molecular magnets and atomic clusters,
display a variety of intrigue physical properties.
It is interesting to compare our theoretical prediction
with experimental results for samples containing
small number of transition-metal dimers of $M=1$, 2 and 3. 
Unfortunately experiments on
samples with such a very small number of dimers, 
have not been reported.
Theoretical and experimental
studies on nanoclusters with changing $M$
could clarify a link between the behavior of the
low-dimensional infinite systems and nanoscale finite-size systems.
The unsettled issues on $T-\beta$ 
and the $N-q$ relations in the current NES
are expected to be resolved by future theoretical and
experimental studies on nanosystems, which
are expected to be one of ideal systems
for a study on the NES.
Our discussion in this study has been confined to the static property
of nanoclusters.
It would be interesting to investigate dynamics of dimers
which has been discussed within the framework of the BGS.

\section*{Acknowledgement}
This work is partly supported by
a Grant-in-Aid for Scientific Research from the Japanese 
Ministry of Education, Culture, Sport, Science 
and Technology.

\vspace{0.5cm}
\appendix*

\section{NES for Heisenberg dimers}
\renewcommand{\theequation}{A\arabic{equation}}
\setcounter{equation}{0}

We consider a cluster containing $M$ Heisenberg dimers,
spin dimers described by the Heisenberg model ($s=1/2$), given by
\begin{eqnarray}
H &=& \sum_{\ell=1}^M H_{\ell}^{(d)}, \\
H_{\ell}^{(d)} &=& -J {\bf s}_1 \cdot {\bf s}_2 
- g \mu_B B (s_{1z}+s_{2z}),
\hspace{1cm}\mbox{($1,2 \in \ell$)}
\end{eqnarray}
where $J$ stands for the exchange interaction,
$g$ (=2) the g-factor, $\mu_B$ the Bohr magneton,
and $B$ an applied magnetic field.
Four eigenvalues for $H_{\ell}^{(d)}$ are given by
\begin{eqnarray}
\epsilon_i &=& -\frac{J}{4}-g \mu_B B m_i,
\hspace{2cm}\mbox{with $m_1=1,\:0,\: -1$ for $i=1, 2, 3$}
\nonumber \\
&=& \frac{3 J}{4}-g \mu_B B m_i.
\hspace{2cm}\mbox{with $m_4=0$ for $i=4$}
\end{eqnarray}
The number of eigenvalues of $H$ becomes $4^M$.

In the BGS the canonical partition function is given by
\begin{eqnarray}
Z_{BG} &=& [Z_{BG}^{(d)}]^M, \\
Z_{BG}^{(d)} &=& {\rm exp}
\left(\frac{\beta J}{4} \right)
\left[1 + 2 {\rm cosh} \left(2 \beta \mu_B B \right) \right] 
+{\rm exp}\left(-\frac{3 \beta J}{4} \right),
\end{eqnarray}
with which thermodynamical quantities are
easily calculated.
The susceptibility is, for example, given by
\cite{Mentrup99,Efremov02}
\begin{eqnarray}
\chi_{BG} &=& M \chi_{BG}^{(d)}, \\
\chi_{BG}^{(d)} &=&
\left( \frac{\mu_B^2}{k_B T} \right) 
\left( \frac{8}{3+ {\rm exp}(-J/k_B T)} \right).
\end{eqnarray}

The calculation of thermodynamical quantities
in the NES for the Heisenberg
model goes parallel
to that discussed in Sec. 2 if we employ
eigenvalues given by Eq. (A3).
For example, by using Eq. (59),
we get the zero-field susceptibility for 
Heisenberg dimers, given by
\begin{eqnarray}
\chi_q 
&=& g^2 \mu_B^2 \left( \frac{q \beta}{c_q} \right)
\: \frac{1}{X_q} \sum_i w_i^{2q-1} m_i^2,
\end{eqnarray}
where a sum $\sum_i$ is performed over $4^M$
eigenvalues [see Eq. (23)].
In the case of $M=1$ (a single dimer), we get
\begin{eqnarray}
\chi_q^{(d)} &=& g^2 \mu_B^2 \left( \frac{q \beta}{c_q} \right)
\left( \frac{2}{X_q} \right)
\left({\rm exp}_q \left[ \left( \frac{\beta}{c_q} \right) 
\left( \frac{J}{4}+E_q \right) \right]
\right)^{2q-1}, \\
X_q &=& 3 \:{\rm exp}_q
\left[ \left( \frac{\beta}{c_q}\right)
\left( \frac{J}{4}+E_q \right) \right]
+{\rm exp}_q
\left[\left( -\frac{\beta}{c_q} \right)
\left( \frac{3J}{4}-E_q \right) \right], \\
E_q &=& \frac{1}{X_q} \{
\left(\frac{-3J}{4} \right) 
\left( {\rm exp}_q \left[ \left( \frac{\beta}{c_q} \right)
\left( \frac{J}{4}+E_q \right) \right] \right)^q \nonumber \\
&+& \left( \frac{3J}{4} \right)
\left( {\rm exp}_q \left[\left(-\frac{\beta}{c_q} \right)
\left (\frac{3J}{4}-E_q \right) \right] \right)^q 
\}.
\end{eqnarray}
In the limit of $q=1$,
Eq. (A9) reduces to $\chi_{BG}^{(d)}$ given by Eq. (A7).

The Curie constant $\Gamma_q$ defined by
$\chi_q=(\mu_B^2/k_B)(\Gamma_q/T)$ for $J=0$ is given by 
\cite{Hasegawa05}
\begin{eqnarray}
\Gamma_q &=& 2 M \:q \:4^{M(q-1)}, 
\hspace{2cm}\mbox{(method A)} \\
&=& 2 M \: q,
\hspace{4cm}\mbox{(method B)} 
\end{eqnarray}
which
are consistent with results obtained 
for Hubbard dimes \cite{Hasegawa05}.
Equations (A12) leads to an anomalously large Curie constant,
which was referred to as {\it dark magnetism} in Ref. 
\cite{Mar00}.

\vspace{0.5cm}
\begin{center}
{$------- Fig. 12 ---------$}
\end{center}
\vspace{0.5cm}

Figure 12(a) and 12(b) show the temperature dependences of
the specific heat and susceptibility, respectively,
of a single Heisenberg dimer ($M=1$) for several $q$ values
for $J < 0$ (antiferromagnetic coupling) 
calculated by the method A (solid curves)
and B (dashed curves).
Figures 12(c) and 12(d) show
the specific heat and susceptibility, respectively, of
two spin dimers ($M=2$) calculated with the
use of $4^2$ eigenvalues.

\vspace{0.5cm}
\begin{center}
{$------- Fig. 13 ---------$}
\end{center}
\vspace{0.5cm}

Figures 13(a) and 13(b) show $C_q$ and $\chi_q$ 
when the size $M$ of a cluster of Heisenberg dimers is changed, with
$q=$ 2.0, 1.5, 1.333 and 1.25 for $M=$ 1, 2, 3 and 4, 
respectively, calculated by the method A:
results for $q=1$ of the BGS
[corresponding to $M=\infty$ in Eq. (3)] are included
for a comparison.
Figs. 13(c) and 13(d) show similar results of $C_q$ and $\chi_q$ 
calculated by the method B.
The $M$ dependence of $C_q$ and $\chi_q$ for Heisenberg dimers 
shown in Fig. 13(a)-13(d) is quite similar to those
shown in Figs. 6(a)-6(d) for Hubbard dimers.



\newpage

\begin{figure}
\caption{
The temperature dependences of the energy $E_q$ for $B=0$ 
with (a) $U/t=0$, (b) 5 and (c) 10 calculated by
the method A,
and those with (d) $U/t=0$, (e) 5 and (f) 10
calculated by the method B:
$q=1.0$ (bold solid curves), 1.1 (dotted curves), 
1.2 (dashed curves), 1.5 (chain curves) 
and 2.0 (solid curves).
}
\label{fig1}
\end{figure}

\begin{figure}
\caption{
The temperature dependences of the entropy $S_q$ for $B=0$
with (a) $U/t=0$, (b) 5 and (c) 10 calculated by
the method A,
and those with (d) $U/t=0$, (e) 5 and (f) 10
calculated by the method B:
$q=1.0$ (bold solid curves), 1.1 (dotted curves), 
1.2 (dashed curves), 1.5 (chain curves) 
and 2.0 (solid curves).
}
\label{fig2}
\end{figure}

\begin{figure}
\caption{
The temperature dependences of the specific heat $C_q$ for $B=0$
with (a) $U/t=0$, (b) 5 and (c) 10 calculated by
the method A,
and those with (d) $U/t=0$, (e) 5 and (f) 10
calculated by the method B:
$q=1.0$ (bold solid curves), 1.1 (dotted curves), 
1.2 (dashed curves), 1.5 (chain curves) 
and 2.0 (solid curves).
}
\label{fig3}
\end{figure}

\begin{figure}
\caption{
The temperature dependences of the susceptibility $\chi_q$
for $B=0$
for (a) $U/t=0$, (b) 5 and (c) 10 calculated by
the method A,
and those for (d) $U/t=0$, (e) 5 and (f) 10
calculated by the method B:
$q=1.0$ (bold solid curves), 1.1 (dotted curves), 
1.2 (dashed curves), 1.5 (chain curves) 
and 2.0 (solid curves).
}
\label{fig4}
\end{figure}

\begin{figure}
\caption{
The temperature dependences of (a) specific heat $C_q$ 
and (b) susceptibility $\chi_q$ (per dimer) 
of Hubbard dimers for $U/t=0$
calculated by the method A, and
those of (c) specific heat $C_q$ 
and (d) susceptibility $\chi_q$
calculated by the method B,
with $M=1$ (bold solid curves),
$M=2$ (chain curves), $M=3$ (dashed curves)
and $M= \infty$ (solid curves).
}
\label{fig5}
\end{figure}

\begin{figure}
\caption{
The temperature dependences of (a) specific heat $C_q$ 
and (b) susceptibility $\chi_q$ (per dimer) 
of Hubbard dimers for $U/t=5$
calculated by the method A, and
those of (c) specific heat $C_q$
and (d) susceptibility $\chi_q$
calculated by the method B, with
$M=1$ (bold solid curves),
$M=2$ (chain curves), $M=3$ (dashed curves)
and $M= \infty$ (solid curves).
}
\label{fig6}
\end{figure}

\begin{figure}
\caption{
(a) $1/M$ dependence of the temperatures of 
$T^*_C$ (circles) and $T^*_{\chi}$ (squares) where 
$C_q$ and $\chi_q$ have the maximum values, respectively.
(b) $1/M$ dependence of the maximum values of 
$C^*_q$ (circles) and $\chi^*_q$ (squares).
Solid and dashed lines denote the results calculated 
by the methods A and B, respectively:
$T^*_{\chi}$ calculated by the method A shown in (a)
is divided by a factor of five.
}
\label{fig7}
\end{figure}

\begin{figure}
\caption{
The magnetization $m_q$ as a function of
the magnetic field $B$
for (a) $U/t=0$, (b) 1 and (c) 10 at $k_B T/t=1.0$:
$q=0.8$ (double-chain curves), 0.9 (chain curves), 
1.0 (bold solid curves),
1.1 (dotted curves), 1.2 (dashed curves), and 1.5 (solid curves)
calculated by the method A.
}
\label{fig8}
\end{figure}

\begin{figure}
\caption{
The magnetic-filed dependence of
the eigenvalues $\epsilon_i$ ($i = 1-6$),
for U/t=5.
}
\label{fig9}
\end{figure}

\begin{figure}
\caption{
The magnetic-filed dependence of
(a) the magnetization $m_q$ 
for $k_B T/t=1.0$ and (b) $k_B T/t=0. 1$,
(c) the susceptibility
for $k_B T/t=1.0$ and (d) $k_B T/t=0. 1$,
(e) the specific heat $\chi_q$ 
for $k_B T/t=1.0$ and (f) $k_B T/t=0. 1$,
of a single Hubbard dimer ($M=1$)
for U/t=5, 
calculated by the method A (solid curves)
method B in the NES (dashed curves), 
and in the BGS (chain curves).
}
\label{fig10}
\end{figure}

\begin{figure}
\caption{
The distributions of $f^B(\beta)$ (solid curves) and
$f^T(\beta)$ (dashed curves) as a function of $\beta$
(see text).
}
\label{fig11}
\end{figure}


\begin{thebibliography}{99}

\bibitem{Tsallis88}C. Tsallis, J. Stat. Phys. {\bf 52}, 479 (1988). 

\bibitem{Tsallis98}C. Tsallis, R. S. Mendes and AA. R. Plastino,
Physica A {\bf 261}, 534 (1998). 

\bibitem{Tsallis04}For a recent review on the NES, see
C. Tsallis, Physica D {\bf 193}, 3 (2004).

\bibitem{NES}Lists of many applications of the nonextensive 
statistics are available
at URL: http://tsallis.cat.cbpf.br/biblio.htm.


\bibitem{Hasegawa05}H. Hasegawa, 
Physica A {\bf 351}, 273 (2005).

\bibitem{Hasegawa05b}H. Hasegawa, 
Proceedings of International Workshop of
CN-Kyoto, March 14-18, 2005
[Prog. Theor. Phys. Suppl. (in press)].

\bibitem{Kakehashi04}Y. Kakehashi, Adv. Phys. {\bf 53}, 497 (2004);
related references therein.

\bibitem{Bader02}S. D. Bader, 
Surf. Sci. {\bf 500}, 172 (2002).

\bibitem{Kach03}H. Kachkachi and D. A. Garanin, 
e-print: cond/mat/0310694.


\bibitem{Luban04}M. Luban, 
J. Magn. Magn. Mat. {\bf 272-276}, e635 (2004);

\bibitem{Fe2}F. Le Gall, F. F. DeBiani, A. Caneschi,
P. Cinelli, A. Cornia, A. C. Fabretti, and D. Gatteschi,
Inorg. Chim. Acta {\bf 262}, 123 (1997); 
A. Lascialfari, F. Tabak, G. L. Abbati, F. Borsa,
M. Corti, and D. Gatteschi,
J. Appl. Phys. {\bf 85}, 4539 (1999).

\bibitem{Mentrup99}D. Mentrup, J. Schnack, and M. Luban,
Physica A {\bf 272}, 153 (1999).
\bibitem{Efremov02}D. V. Efremov and R. A. Klemm,
Phys. Rev. B {\bf 66}, 174427 (2002);
\bibitem{Dai03}D. Dai and M. Whangbo,
J. Chem. Phys. {\bf 118}, 29 (2003).

\bibitem{V2}Y. Furukawa, A. Iwai, K. Kumagai, and
A. Yabubovsky,
J. Phys. Soc. Jpn. {\bf 65}, 2393 (1996);
D. A. Tennant, S. E. Nagler, A. W. Garrett,
T. Barnes, and C. C. Torardi,
Phys. Rev. Lett. {\bf 78}, 4998 (1997);
A. W. Garrett, S. E. Nagler, D. A. Tennant,
B. C. Sales, and T. Barnes,
Phys. Rev. Lett. {\bf 79}, 745 (1997).

\bibitem{Cr2}M. S. Bailey, M. N. Obrovac, E. Baillet,
T. K. Reynolds and F. J. DiSalvo,
Inorg. Chem. {\bf 42}, 5572 (2003);
J. Glerup, P. A. Goodson, D. J. Hodgson,
M. A. Masood, and K. Michelsen,
Inorganica {\bf 358}, 295 (2005).

\bibitem{Co2}U. Beckmann and S. Brooker,
Coordination Chemistry {\bf 245}, 17 (2003);
N. D. Lazarov, V. Spasojevic, V. Kusigerski,
V. M. Mati\'{c} and M. Mili\'{c},
J. Magn. Magn. Matt. {\bf 272-276}, 1065 (2004).

\bibitem{Ni2}S. K. Dey, M. S. E. Fallah, J. Ribas,
T. Matsushita, V. Gramlich and S. Mitra,
Inorganica Chmica {\bf 357}, 1517 (2004).

\bibitem{Cu2}A. Zheludev, G. Shirane, Y. Sasago, M. Hase,
and K. Uchinokura,
Phys. Rev. B {\bf 53}, 11642 (1996).

\bibitem{Bernstein74}U. Bernstein and P. Pincus, 
Phys. Rev. B {\bf 10}, 3626 (1974).

\bibitem{Wilk00}G. Wilk and Z. Wlodarczyk,
Phys. Rev. Lett. {\bf 84}, 2770 (2000).

\bibitem{Beck02}C. Beck, 
Europhys. Lett. {\bf 57}, 329 (2002).

\bibitem{Raja04}A. K. Rajagopal, C. S. Pande, and S. Abe,
e-print cond-mat/0403738.

\bibitem{Abe01}S. Abe, S. Mart\'{i}nez, F. Pennini
and A. Plastino,
Phys. Lett. A {\bf 281}, 126 (2001).

\bibitem{Abe99}S. Abe, Phys. Lett. A {\bf 263}, 424 (1999):
{\it ibid.} {\bf 267}, 456 (2000) (erratum).

\bibitem{Suezaki72}Y. Suezaki, Phys. Lett. {\bf 38A}, 293 (1972).

\bibitem{Shiba72}H. Shiba and P. A. Pincus, 
Phys. Rev. B {\bf 5}, 1966 (1972).

\bibitem{Kuzmenko04}N. K. Kuzmenkoand V. M. Mikhajlov,
e-print cond-mat/0401468.


\bibitem{Mn12}
D. Gatteschia and R. Sessoli,
J. Magn. Magn. Mat. {\bf272}, 272 (2004);
related references therein.

\bibitem{Touchette02}H. Touchette, e-print cond-mat/0212301.


\bibitem{Beck04}C. Beck and E. G. D. Cohen,
e-print cond-mat/0205097;
H. Touchette and C. Beck,
e-print cond-mat/0408091.

\bibitem{Note9}The Curie constant of
the BGS susceptibility given by Eq. (61)
reduces to
$\Gamma_{BG}= [8/(4+2 {\rm cosh}(\beta t)]$ in the limit
of $U =0$, to
$\Gamma_{BG} = [8/(3+e^{4\beta t^2/U}+2 e^{-\beta U})]$ in the limit
of $t/U \rightarrow 0$, and to
$\Gamma_{BG}= [8/(4+2 e^{-\beta U})]$ in the limit
of $t = 0$.

\bibitem{Suyari05}H. Suyari, e-print cond-mat/0502298.

\bibitem{Wada05}T. Wada and A. M. Scarfone,
e-print cond-mat/0502394.

\bibitem{Abe05}S. Abe,
e-print cond-mat/0504036.



\bibitem{Mar00}S. Martinez, F. Pennini, and A. Plastino,
Physica A {\bf 282}, 193 (2000).




\end{thebibliography}
\end{document}